\documentclass[a4paper]{article}
\usepackage{epsfig,amsmath,amsfonts,amssymb,setspace,multirow,textcomp}
\usepackage[T1]{fontenc}
\makeatletter
\renewcommand{\@oddhead}{\hfil}
\renewcommand{\@evenfoot}{\hfil \thepage \hfil}
\renewcommand{\@oddfoot}{\hfil \thepage \hfil}
\makeatother

\renewenvironment{thebibliography}[1]{\begin{oldthebibliography}{#1}\setlength{\parskip}{0ex}\setlength{\itemsep}{0ex}}{\end{oldthebibliography}}
\begin{document}
\fontsize{11}{11}\selectfont % the font size cannot be changed in any case!
%  insert your title, authors information and text instead of the one provided below
\title{Constraints on the dark energy with barotropic equation of state: assessing the importance of different observations}
\author{\textsl{O.~Sergijenko}}
\date{\vspace*{-6ex}}
\maketitle
\begin{center} {\small Astronomical Observatory, Taras Shevchenko National University of Kyiv, Observatorna str. 3, Kyiv, 04053, Ukraine\\
{\tt olga.sergijenko.astro@gmail.com}}
\end{center}

\begin{abstract}
For dynamical dark energy with the barotropic equation of state we determine the mean values of parameters and their confidence ranges together with other cosmological parameters on the basis of different combined datasets. The used observations include Planck data on CMB temperature anisotropy, E-mode polarization and lensing, BICEP2/Keck Array data on B-mode polarization, BAO from SDSS and 6dFGS, power spectrum of galaxies from WiggleZ, weak lensing from CFHTLenS and SN Ia data from the JLA compilation. We find that all but one mean models are phantom, mean values of the equation of state parameter at current epoch are close to $-1$ and constraints on the adiabatic sound speed of dark energy are weak. We investigate the effect of CMB polarization data on the dark energy parameters estimation. We discuss also which type of data on the large scale structure of the Universe allows to determine the dark energy parameters most precisely.\\[1ex]
\end{abstract}

\section*{\sc introduction}
\indent \indent The final analysis of data obtained by the Planck satellite \cite{planck2018} shows that the dark energy at current epoch is close to the cosmological constant: $w_0=-1.028\pm0.032$. However, the $\Lambda$ term faces numerous interpretational problems.

One of the simplest alternatives to $\Lambda$ is the dynamical dark energy in form of the minimally coupled classical scalar field. In previous papers (e. g. \cite{novosyadlyj2011,novosyadlyj2013,novosyadlyj2014}) we have introduced and studied in detail such field with the barotropic equation of state. This model involves both quintessential and phantom subclasses, its equation of state parameter and energy density have the analytical form applicable at any time during the past history of the Universe.

The goal of this paper is to obtain new constraints on parameters of the dynamical dark energy with barotropic equation of state from more recent data than used in \cite{novosyadlyj2014} and to investigate which data are most useful for tightening the constraints on dark energy parameters.

\section*{\sc cosmological model}
\indent \indent We suppose that the Universe is spatially flat, homogeneous and isotropic with Friedmann-Robertson-Walker (FRW) metric of 4-space
\begin{eqnarray}
ds^2=g_{ij} dx^i dx^j =a^2(\eta)(d\eta^2-\delta_{\alpha\beta} dx^{\alpha}dx^{\beta})\label{metr}
\end{eqnarray}
(here $i,j=0,1,2,3$, $\alpha,\beta=1,2,3$, $a$ is the scale factor, $\eta$ is the conformal time and $c=1$). It is filled with photons, neutrinos, baryons, cold dark matter and dark energy. For neutrinos we apply the minimal-mass normal hierarchy of masses: a single massive eigenstate with $m_{\nu}=0.06$ eV.

The dark energy is assumed to be the minimally coupled classical scalar field with barotropic equation of state, which is described in detail in \cite{novosyadlyj2014}. It has the equation of state (EoS) parameter:
\begin{equation}
w_{de}=\frac{p_{de}}{\rho_{de}}=\frac{(1+c_a^2)(1+w_0)}{1+w_0-(w_0-c_a^2)a^{3(1+c_a^2)}}-1\label{w}
\end{equation}
(here and below $0$ denotes the values at current time and $c_a^2=const$ is the adiabatic sound speed of dark energy, which has the meaning of EoS parameter at the Big Bang if $c_a^2>-1$ and at the infinite time if $c_a^2<-1$, the asymptotic values in the opposite time directions are $-1$, the case $c_a^2=-1$ corresponds to the cosmological constant), \\
energy density:
\begin{equation}
\rho_{de}=\rho_{de}^{(0)}\frac{(1+w_0)a^{-3(1+c_a^2)}+c_a^2-w_0}{1+c_a^2}\label{rho}
\end{equation}
and effective sound speed $c_s^2=1$.

We exclude from consideration the models with $w_0<-1$ and $c_a^2>w_0$, because for such values in the past the EoS parameter had discontinuity of the second kind and $\rho_{de}$ changed the sign.

In the case of flat 3-space the dark energy model has 2 free parameters: $w_0$ and $c_a^2$ (the third one, $\Omega_{de}=\rho_{de}^{(0)}/\rho_{cr}^{(0)}$, where $\rho_{cr}$ is the crytical density, is determined as $\Omega_{de}=1-\Omega_b-\Omega_{cdm}$, where $\Omega_b=\rho_b^{(0)}/\rho_{cr}^{(0)}$ and $\Omega_{cdm}=\rho_{cdm}^{(0)}/\rho_{cr}^{(0)}$).

We also assume the slow-roll inflation.

\section*{\sc method and data}

\indent \indent We determine the dark energy parameters $w_0$ and $c_a^2$ jointly with other cosmological ones: $\Omega_bh^2$, $\Omega_{cdm}h^2$, $h\equiv H_0/100$\,km/(s$\cdot$Mpc) (here $H_0$ is the Hubble constant), the amplitude of primordial power spectrum $A_s$, the scalar spectral index $n_s$, the optical depth to reionization $\tau_{rei}$ and the tensor-to-scalar ratio $r\equiv A_t/A_s$).

To estimate these parameters we use the Monte Carlo Markov chain (MCMC) method implemented in the CosmoMC code \cite{cosmomc}. The CAMB code is used to compute the theory predictions \cite{camb}. The dark energy perturbations are treated within the parametrized post-Friedmann (PPF) framework \cite{fang2008}. When necessary we use the nonlinear corrections by HALOFIT adopted for the studied type of dark energy.

Combined datasets include the following CMB observations:
\begin{itemize}
\item Planck: data on the TT, TE, EE spectra and lensing \cite{planck2015} (the usefulness of EE spectra to distinguish between different subclasses of the dark energy models with barotropic EoS was forecasted in \cite{novosyadlyj2011,novosyadlyj2013});
\item BICEP2/Keck Array+Planck: data on the B-mode polarization in 1 frequency band 150 GHz \cite{bkp} (hereafter BKP);
\item BICEP2/Keck Array: data on the B-mode polarization in 2 frequency bands 150 GHz and 95 GHz \cite{bk} (hereafter BK).
\end{itemize}

We use 3 types of the data on large scale structure of the Universe (LSS):
\begin{itemize}
\item BAO 6dFGS \cite{bao6df} and SDSS MGS \cite{baomgs} (hereafter BAO),
\item power spectrum of galaxies from WiggleZ \cite{wigglez} or
\item weak lensing from CFHTLenS \cite{cfhtlens}.
\end{itemize}

The Supernovae Ia luminosity distances and redshifts are taken from the JLA compilation \cite{jla}.

We do not discuss the tension between values of $H_0$ inferred from CMB \cite{planck2018} and obtained from direct measurements (e. g. \cite{riess2018}). We take into account the prior on $H_0$ from the reanalysis \cite{hst} which is in better agreement with CMB.

We use 12 different combined datasets in total.

For the dark energy parameters $w_0$ and $c_a^2$ we apply flat priors with the ranges of values [-2,-0.33] and [-2,0] correspondingly.

Each MCMC run has 8 chains converged to $R<0.01$.

\begin{figure}[!h]
\centering
\begin{minipage}[t]{.49\linewidth}
\centering
\epsfig{file = 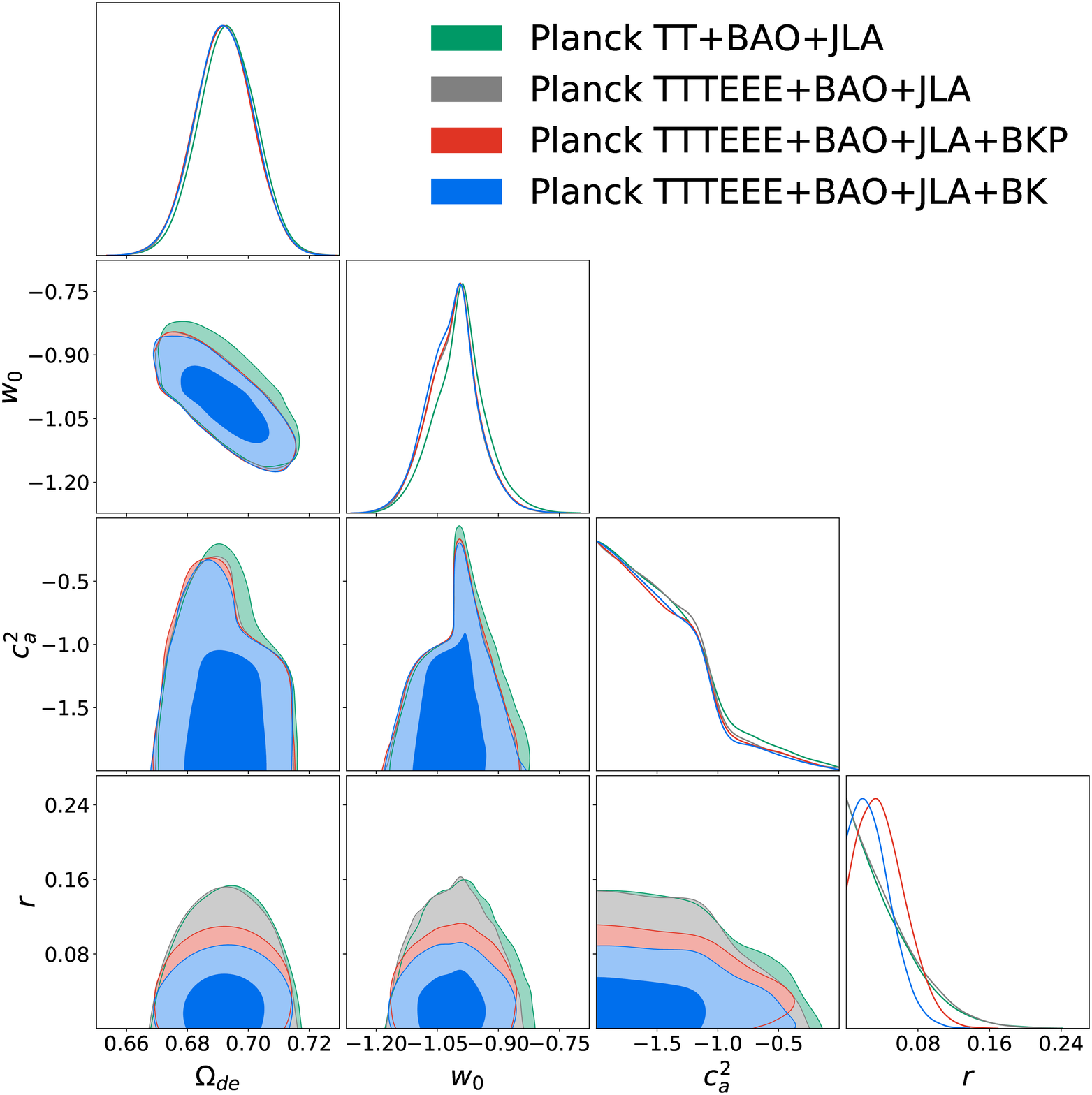,width = \linewidth}
\caption{1D marginalized posteriors for $\Omega_{de}$, $w_0$, $c_a^2$, $r$ and $1\sigma$, $2\sigma$ confidence contours from the 2D marginalized posterior distributions for the combined datasets including BAO.}\label{fig1}
\end{minipage}
\hfill
\begin{minipage}[t]{.49\linewidth}
\centering
\epsfig{file = 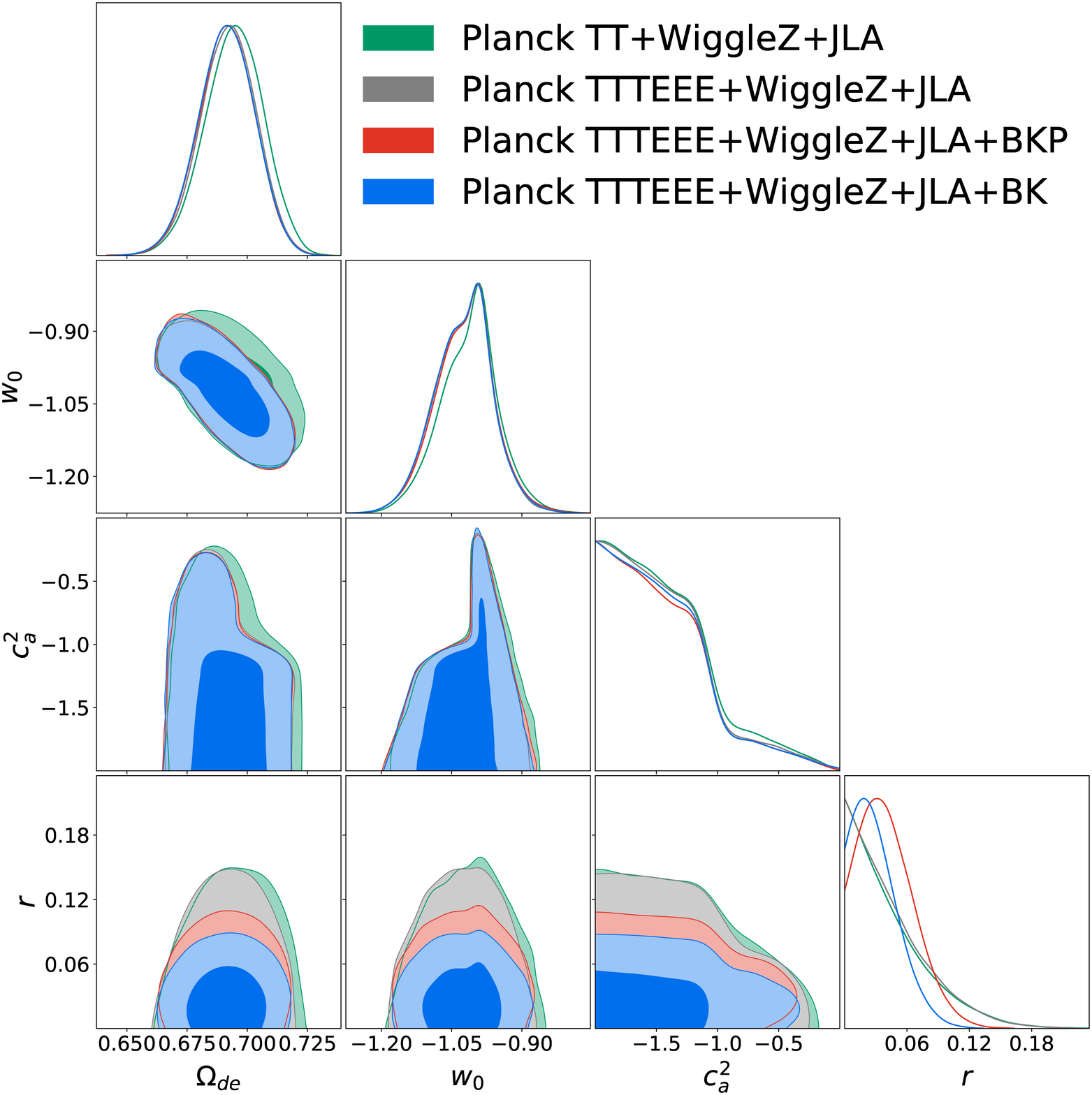,width = \linewidth}
\caption{1D marginalized posteriors for $\Omega_{de}$, $w_0$, $c_a^2$, $r$ and $1\sigma$, $2\sigma$ confidence contours from the 2D marginalized posterior distributions for the combined datasets including the power spectrum of galaxies.}\label{fig2}
\end{minipage}
\end{figure}

\begin{figure}[!h]
\centering
\begin{minipage}[t]{.49\linewidth}
\centering
\epsfig{file = 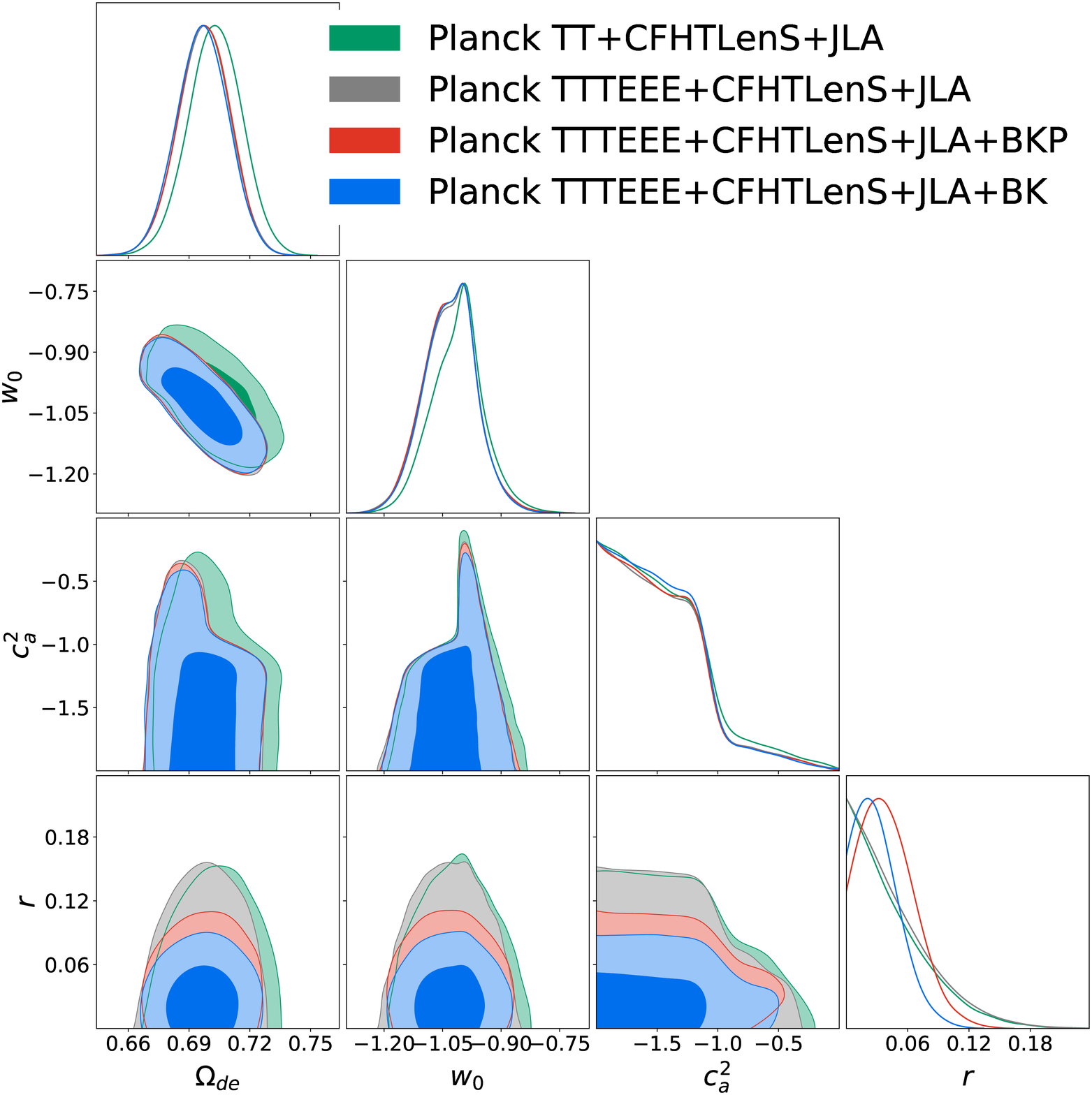,width = \linewidth}
\caption{1D marginalized posteriors for $\Omega_{de}$, $w_0$, $c_a^2$, $r$ and $1\sigma$, $2\sigma$ confidence contours from the 2D marginalized posterior distributions for the combined datasets including weak lensing.}\label{fig3}
\end{minipage}
\hfill
\begin{minipage}[t]{.49\linewidth}
\centering
\epsfig{file = 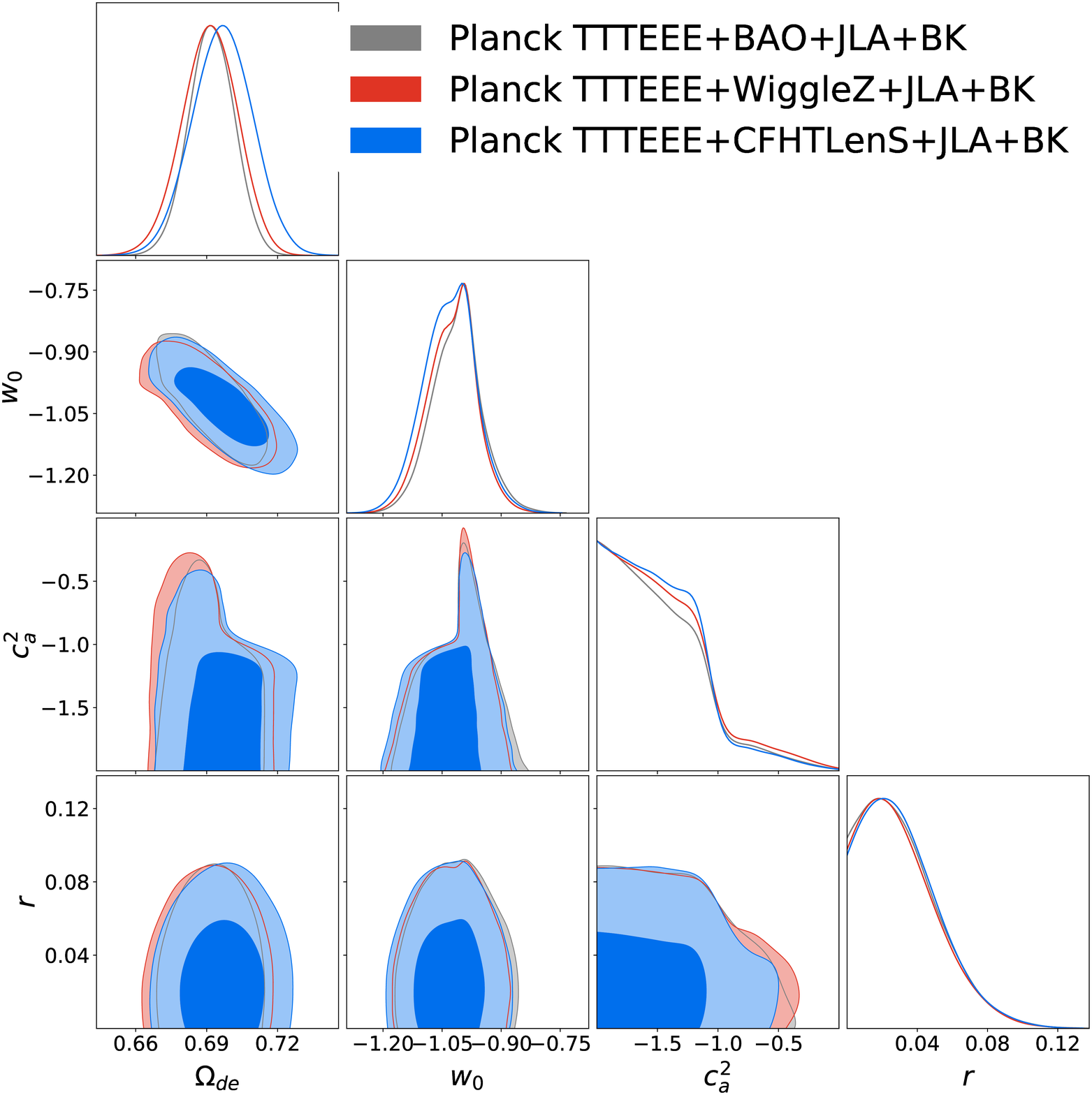,width = \linewidth}
\caption{1D marginalized posteriors for $\Omega_{de}$, $w_0$, $c_a^2$, $r$ and $1\sigma$, $2\sigma$ confidence contours from the 2D marginalized posterior distributions for the combined datasets including different types of the LSS data.}\label{fig4}
\end{minipage}
\end{figure}

\begin{table}
 \centering
 \caption{The mean values, 1$\sigma$ and 2$\sigma$ confidence limits for cosmological parameters obtained from the observational datasets including CMB, BAO and JLA.}\label{tab1}
 \vspace*{1ex}
  \begin{tabular}{ccccc}
  \hline
  Parameters&Planck TT&Planck TTTEEE&Planck TTTEEE+BKP&Planck TTTEEE+BK\\
  &mean$\pm1\sigma\pm2\sigma$&mean$\pm1\sigma\pm2\sigma$&mean$\pm1\sigma\pm2\sigma$&mean$\pm1\sigma\pm2\sigma$\\
  \hline
$\Omega_{de}$ & 0.693$_{-  0.009}^{+  0.010}$ $_{-  0.019}^{+  0.018}$&0.692$_{-  0.009}^{+  0.009}$ $_{-  0.019}^{+  0.019}$&0.692$_{-  0.010}^{+  0.010}$ $_{-  0.019}^{+  0.019}$&0.692$_{-  0.010}^{+  0.010}$ $_{-  0.019}^{+  0.019}$\medskip\\
$w_0$ &-0.994$_{-  0.064}^{+  0.060}$ $_{-  0.133}^{+  0.139}$&-1.010$_{-  0.062}^{+  0.057}$ $_{-  0.125}^{+  0.128}$&-1.010$_{-  0.062}^{+  0.058}$ $_{-  0.129}^{+  0.130}$&-1.013$_{-  0.062}^{+  0.058}$ $_{-  0.127}^{+  0.130}$\medskip\\
$c_a^2$ &-1.448$_{-  0.552}^{+  0.143}$ $_{-  0.552}^{+  0.808}$&-1.467$_{-  0.533}^{+  0.147}$ $_{-  0.533}^{+  0.739}$&-1.474$_{-  0.526}^{+  0.148}$ $_{-  0.526}^{+  0.755}$&-1.485$_{-  0.515}^{+  0.143}$ $_{-  0.515}^{+  0.731}$\medskip\\
$r$ & 0.045$_{-  0.045}^{+  0.010}$ $_{-  0.045}^{+  0.075}$&0.046$_{-  0.046}^{+  0.010}$ $_{-  0.046}^{+  0.074}$&0.042$_{-  0.036}^{+  0.015}$ $_{-  0.042}^{+  0.048}$&0.032$_{-  0.032}^{+  0.008}$ $_{-  0.032}^{+  0.041}$\medskip\\
$10\Omega_bh^2$ & 0.223$_{-  0.002}^{+  0.002}$ $_{-  0.004}^{+  0.004}$&0.223$_{-  0.001}^{+  0.001}$ $_{-  0.003}^{+  0.003}$&0.223$_{-  0.001}^{+  0.001}$ $_{-  0.003}^{+  0.003}$&0.223$_{-  0.001}^{+  0.001}$ $_{-  0.003}^{+  0.003}$\medskip\\
$\Omega_{cdm}h^2$ & 0.118$_{-  0.001}^{+  0.001}$ $_{-  0.003}^{+  0.003}$&0.119$_{-  0.001}^{+  0.001}$ $_{-  0.002}^{+  0.002}$&0.119$_{-  0.001}^{+  0.001}$ $_{-  0.002}^{+  0.002}$&0.119$_{-  0.001}^{+  0.001}$ $_{-  0.002}^{+  0.002}$\medskip\\
$h$ & 0.678$_{-  0.011}^{+  0.010}$ $_{-  0.020}^{+  0.021}$&0.679$_{-  0.011}^{+  0.010}$ $_{-  0.020}^{+  0.021}$&0.679$_{-  0.011}^{+  0.010}$ $_{-  0.020}^{+  0.021}$&0.679$_{-  0.011}^{+  0.010}$ $_{-  0.020}^{+  0.021}$\medskip\\
$n_s$ & 0.969$_{-  0.005}^{+  0.005}$ $_{-  0.010}^{+  0.010}$&0.967$_{-  0.004}^{+  0.004}$ $_{-  0.008}^{+  0.009}$&0.967$_{-  0.004}^{+  0.004}$ $_{-  0.008}^{+  0.008}$&0.967$_{-  0.004}^{+  0.004}$ $_{-  0.008}^{+  0.008}$\medskip\\
$\log(10^{10}A_s)$ & 3.066$_{-  0.026}^{+  0.026}$ $_{-  0.052}^{+  0.052}$&3.061$_{-  0.024}^{+  0.024}$ $_{-  0.048}^{+  0.048}$&3.063$_{-  0.024}^{+  0.024}$ $_{-  0.047}^{+  0.048}$&3.064$_{-  0.024}^{+  0.024}$ $_{-  0.047}^{+  0.047}$\medskip\\
$\tau_{rei}$ & 0.068$_{-  0.014}^{+  0.014}$ $_{-  0.028}^{+  0.029}$&0.065$_{-  0.013}^{+  0.013}$ $_{-  0.026}^{+  0.026}$& 0.065$_{-  0.013}^{+  0.013}$ $_{-  0.025}^{+  0.026}$&0.066$_{-  0.013}^{+  0.013}$ $_{-  0.025}^{+  0.025}$\medskip\\
\hline
 \end{tabular}
\end{table}
 
\begin{table}
 \centering
 \caption{The mean values, 1$\sigma$ and 2$\sigma$ confidence limits for cosmological parameters obtained from the observational datasets including CMB, WiggleZ and JLA.}\label{tab2}
 \vspace*{1ex}
  \begin{tabular}{ccccc}
  \hline
  Parameters&Planck TT&Planck TTTEEE&Planck TTTEEE+BKP&Planck TTTEEE+BK \cite{sergijenko2018}\\
  &mean$\pm1\sigma\pm2\sigma$&mean$\pm1\sigma\pm2\sigma$&mean$\pm1\sigma\pm2\sigma$&mean$\pm1\sigma\pm2\sigma$\\
  \hline
$\Omega_{de}$ & 0.694$_{-  0.011}^{+  0.013}$ $_{-  0.025}^{+  0.023}$&0.692$_{-  0.011}^{+  0.012}$ $_{-  0.024}^{+  0.022}$&0.691$_{-  0.012}^{+  0.012}$ $_{-  0.023}^{+  0.022}$&0.691$_{-  0.012}^{+  0.012}$ $_{-  0.024}^{+  0.022}$\medskip\\
$w_0$ &-1.012$_{-  0.060}^{+  0.060}$ $_{-  0.130}^{+  0.125}$&-1.025$_{-  0.058}^{+  0.062}$ $_{-  0.126}^{+  0.119}$&-1.022$_{-  0.058}^{+  0.061}$ $_{-  0.127}^{+  0.123}$&-1.024$_{-  0.058}^{+  0.062}$ $_{-  0.125}^{+  0.120}$\medskip\\
$c_a^2$ &-1.438$_{-  0.562}^{+  0.150}$ $_{-  0.562}^{+  0.815}$&-1.456$_{-  0.544}^{+  0.146}$ $_{-  0.544}^{+  0.791}$&-1.463$_{-  0.537}^{+  0.147}$ $_{-  0.537}^{+  0.778}$&-1.460$_{-  0.540}^{+  0.145}$ $_{-  0.540}^{+  0.781}$\medskip\\
$r$ & 0.045$_{-  0.045}^{+  0.010}$ $_{-  0.045}^{+  0.075}$&0.045$_{-  0.045}^{+  0.010}$ $_{-  0.045}^{+  0.073}$&0.042$_{-  0.035}^{+  0.016}$ $_{-  0.042}^{+  0.048}$&0.031$_{-  0.031}^{+  0.008}$ $_{-  0.031}^{+  0.041}$\medskip\\
$10\Omega_bh^2$ & 0.223$_{-  0.002}^{+  0.002}$ $_{-  0.004}^{+  0.004}$&0.223$_{-  0.002}^{+  0.002}$ $_{-  0.003}^{+  0.003}$&0.222$_{-  0.002}^{+  0.002}$ $_{-  0.003}^{+  0.003}$&0.222$_{-  0.002}^{+  0.002}$ $_{-  0.003}^{+  0.003}$\medskip\\
$\Omega_{cdm}h^2$ & 0.118$_{-  0.002}^{+  0.002}$ $_{-  0.004}^{+  0.004}$&0.119$_{-  0.001}^{+  0.001}$ $_{-  0.003}^{+  0.003}$&0.119$_{-  0.001}^{+  0.001}$ $_{-  0.003}^{+  0.003}$&0.119$_{-  0.001}^{+  0.001}$ $_{-  0.003}^{+  0.003}$\medskip\\
$h$ & 0.681$_{-  0.012}^{+  0.012}$ $_{-  0.023}^{+  0.024}$&0.680$_{-  0.012}^{+  0.012}$ $_{-  0.023}^{+  0.024}$&0.679$_{-  0.012}^{+  0.012}$ $_{-  0.023}^{+  0.024}$&0.679$_{-  0.012}^{+  0.012}$ $_{-  0.023}^{+  0.024}$\medskip\\
$n_s$ & 0.969$_{-  0.006}^{+  0.006}$ $_{-  0.011}^{+  0.011}$&0.966$_{-  0.005}^{+  0.005}$ $_{-  0.009}^{+  0.009}$&0.966$_{-  0.005}^{+  0.005}$ $_{-  0.009}^{+  0.009}$&0.966$_{-  0.005}^{+  0.005}$ $_{-  0.009}^{+  0.009}$\medskip\\
$\log(10^{10}A_s)$ & 3.062$_{-  0.028}^{+  0.029}$ $_{-  0.057}^{+  0.057}$&3.056$_{-  0.025}^{+  0.025}$ $_{-  0.050}^{+  0.051}$&3.057$_{-  0.025}^{+  0.025}$ $_{-  0.049}^{+  0.050}$&3.059$_{-  0.025}^{+  0.025}$ $_{-  0.049}^{+  0.049}$\medskip\\
$\tau_{rei}$ & 0.066$_{-  0.016}^{+  0.016}$ $_{-  0.031}^{+  0.031}$&0.062$_{-  0.014}^{+  0.014}$ $_{-  0.027}^{+  0.028}$&0.062$_{-  0.014}^{+  0.014}$ $_{-  0.027}^{+  0.027}$&0.063$_{-  0.013}^{+  0.014}$ $_{-  0.027}^{+  0.027}$\medskip\\
\hline
 \end{tabular}
\end{table}

\begin{table}
 \centering
 \caption{The mean values, 1$\sigma$ and 2$\sigma$ confidence limits for cosmological parameters obtained from the observational datasets including CMB, CFHTLenS and JLA.}\label{tab3}
 \vspace*{1ex}
  \begin{tabular}{ccccc}
  \hline
  Parameters&Planck TT&Planck TTTEEE&Planck TTTEEE+BKP&Planck TTTEEE+BK\\
  &mean$\pm1\sigma\pm2\sigma$&mean$\pm1\sigma\pm2\sigma$&mean$\pm1\sigma\pm2\sigma$&mean$\pm1\sigma\pm2\sigma$\\
  \hline
$\Omega_{de}$ & 0.703$_{-  0.013}^{+  0.013}$ $_{-  0.027}^{+  0.026}$&0.697$_{-  0.013}^{+  0.013}$ $_{-  0.026}^{+  0.025}$&0.697$_{-  0.013}^{+  0.013}$ $_{-  0.026}^{+  0.024}$&0.697$_{-  0.013}^{+  0.013}$ $_{-  0.025}^{+  0.025}$\medskip\\
$w_0$ &-1.011$_{-  0.068}^{+  0.062}$ $_{-  0.136}^{+  0.140}$&-1.030$_{-  0.064}^{+  0.066}$ $_{-  0.136}^{+  0.134}$&-1.031$_{-  0.065}^{+  0.067}$ $_{-  0.133}^{+  0.137}$&-1.030$_{-  0.064}^{+  0.066}$ $_{-  0.132}^{+  0.131}$\medskip\\
$c_a^2$ &-1.456$_{-  0.544}^{+  0.151}$ $_{-  0.544}^{+  0.768}$&-1.478$_{-  0.522}^{+  0.150}$ $_{-  0.522}^{+  0.709}$&-1.482$_{-  0.518}^{+  0.151}$ $_{-  0.518}^{+  0.687}$&-1.482$_{-  0.518}^{+  0.151}$ $_{-  0.518}^{+  0.661}$\medskip\\
$r$ & 0.046$_{-  0.046}^{+  0.010}$ $_{-  0.046}^{+  0.074}$&0.047$_{-  0.047}^{+  0.011}$ $_{-  0.047}^{+  0.076}$&0.042$_{-  0.036}^{+  0.016}$ $_{-  0.042}^{+  0.048}$&0.032$_{-  0.032}^{+  0.008}$ $_{-  0.032}^{+  0.040}$\medskip\\
$10\Omega_bh^2$ & 0.224$_{-  0.002}^{+  0.002}$ $_{-  0.004}^{+  0.004}$&0.223$_{-  0.002}^{+  0.002}$ $_{-  0.003}^{+  0.003}$&0.223$_{-  0.002}^{+  0.002}$ $_{-  0.003}^{+  0.003}$&0.223$_{-  0.002}^{+  0.002}$ $_{-  0.003}^{+  0.003}$\medskip\\
$\Omega_{cdm}h^2$ & 0.117$_{-  0.002}^{+  0.002}$ $_{-  0.004}^{+  0.004}$&0.119$_{-  0.001}^{+  0.001}$ $_{-  0.003}^{+  0.003}$&0.119$_{-  0.001}^{+  0.001}$ $_{-  0.003}^{+  0.003}$&0.119$_{-  0.001}^{+  0.001}$ $_{-  0.003}^{+  0.003}$\medskip\\
$h$ & 0.687$_{-  0.014}^{+  0.013}$ $_{-  0.026}^{+  0.028}$&0.684$_{-  0.014}^{+  0.013}$ $_{-  0.026}^{+  0.028}$&0.684$_{-  0.013}^{+  0.014}$ $_{-  0.026}^{+  0.027}$&0.684$_{-  0.014}^{+  0.013}$ $_{-  0.025}^{+  0.028}$\medskip\\
$n_s$ & 0.972$_{-  0.006}^{+  0.006}$ $_{-  0.011}^{+  0.012}$&0.968$_{-  0.005}^{+  0.005}$ $_{-  0.009}^{+  0.009}$&0.968$_{-  0.005}^{+  0.005}$ $_{-  0.009}^{+  0.009}$&0.967$_{-  0.005}^{+  0.005}$ $_{-  0.009}^{+  0.009}$\medskip\\
$\log(10^{10}A_s)$ & 3.072$_{-  0.030}^{+  0.030}$ $_{-  0.058}^{+  0.060}$&3.058$_{-  0.025}^{+  0.025}$ $_{-  0.050}^{+  0.050}$&3.060$_{-  0.025}^{+  0.025}$ $_{-  0.050}^{+  0.050}$&3.061$_{-  0.025}^{+  0.025}$ $_{-  0.049}^{+  0.049}$\medskip\\
$\tau_{rei}$ & 0.072$_{-  0.016}^{+  0.016}$ $_{-  0.032}^{+  0.033}$&0.064$_{-  0.014}^{+  0.014}$ $_{-  0.027}^{+  0.027}$&0.064$_{-  0.014}^{+  0.014}$ $_{-  0.027}^{+  0.027}$&0.065$_{-  0.014}^{+  0.013}$ $_{-  0.027}^{+  0.027}$\medskip\\
\hline
 \end{tabular}
\end{table}
 
\section*{\sc results and discussion}
\indent \indent The results are presented in Figures \ref{fig1}-\ref{fig4} and Tables \ref{tab1}-\ref{tab3}.

The mean values of $w_0$ from Tables \ref{tab1}-\ref{tab3} are much closer to $-1$ than in \cite{novosyadlyj2014,vavilova2015} (where the 2013 year data from Planck were used) for all combined datasets, making the constraints on $c_a^2$ significantly weaker (indeed, as it can be seen in Fig. \ref{fig1}-\ref{fig3}, the posteriors for $c_a^2$ are cut from below by the prior). Among the mean values of $w_0$ only the one obtained from Planck TT+BAO+JLA is larger than $-1$ so the mean model is quintessential. All other mean models are phantom. In all cases the mean value of $c_a^2$ is smaller than $-1$ meaning that in the past the EoS parameter evolved from $-1$ to the current value $w_0$. The models with parameters on upper $1\sigma$ and $2\sigma$ confidence limits are quintessential, those with parameters on lower $1\sigma$ and $2\sigma$ confidence limits are phantom. This is in agreement with results obtained in \cite{sergijenko2017} for the cosmological model with free tensor spectral index $n_t$.

As we see in Fig. \ref{fig1}-\ref{fig3}, the inclusion of data on CMB polarization (either only E or both E and B modes) allows to narrow somewhat the confidence contours for dark energy parameters $w_0$ and $c_a^2$ (this effect is most visible for the datasets including CFHTLenS) and to narrow sufficiently the confidence contours for $\Omega_{de}$.

While the inclusion of E-mode polarization has small effect on the upper limits of tensor-to-scalar ratio $r$ (Fig. \ref{fig1}-\ref{fig3}), the use of BKP and BK data on B-mode polarization reduces them significantly. The upper $2\sigma$ limit for $r$ decreases from 0.12 to 0.073 when the BAO data are used, from 0.12 to 0.072 in case of the WiggleZ data and from 0.12 to 0.072 for CFHTLenS.

In Fig. \ref{fig4} we see that there is some tension between the constraints obtained from datasets including different types of the LSS data, especially for $\Omega_{de}$, but all determinations are within $1\sigma$ limits. BAO data allow the most precise determination for $\Omega_{de}$ (from Table \ref{tab1} the width of $2\sigma$ confidence range is 0.038), WiggleZ for $w_0$ (from Table \ref{tab2} the width of $2\sigma$ confidence range is 0.245) and CFHTLenS for $c_a^2$ (from Table \ref{tab3} the width of $2\sigma$ confidence range is 1.179).

\section*{\sc conclusions}
\indent \indent For 12 combined datasets we determine the mean values of cosmological parameters and their confidence ranges for model with the dark energy with barotropic equation of state. For the mean model all but one combined datasets prefer phantom over quintessence. Evolution of the mean EoS parameter from $-1$ in the early Universe to $w_0$ at current epoch is slow (both values are close) and the constraints on $c_a^2$ are weak. Already at $1\sigma$ level the dark energy models with parameters at the lower confidence limits are phantom and those with parameters at the upper confidence limits are quintessence. Both E and B modes of CMB polarization allow to narrow confidence ranges for the values of dark energy parameters comparing to only CMB TT spectrum and therefore should be included into combined datasets. No type of the LSS data can be favored since the most precise determination for $\Omega_{de}$ is obtained from the combined datasets including BAO, for $w_0$ from the combined datasets including WiggleZ and for $c_a^2$ from the combined datasets including CFHTLenS.

\section*{\sc acknowledgement}
\indent \indent This work was partially supported by the project of Ministry of Education and Science of Ukraine (state registration number 0116U001544). Author also acknowledges the usage of CAMB and CosmoMC packages.

\end{document}